\documentclass[10pt,journal,doublecolumn]{IEEEtran}

\usepackage{color,array,amsthm}
\usepackage{graphicx}

\ifCLASSINFOpdf

\else

\fi
\usepackage{amsmath}
\usepackage{graphicx}
\usepackage{subfig}
\usepackage{diagbox}
\usepackage{caption}
\usepackage{epstopdf}
\usepackage{cite}
\usepackage{amsfonts}
\usepackage{amssymb}
\usepackage{xcolor}
\usepackage{textcomp}
\usepackage{dsfont}
\usepackage{mathrsfs}%
\usepackage{xcolor}%
\usepackage{textcomp}%
\usepackage{manyfoot}%
\usepackage{booktabs}%
\usepackage{algorithm}%
\usepackage{algorithmicx}%
\usepackage{algpseudocode}%
\usepackage{listings}
\usepackage{tikz}
\usetikzlibrary{matrix}
\usepackage{hyperref}
\usepackage{url}
\newtheorem{fact}{{{Fact}}}
\newtheorem{example}{{{Example}}}
\newtheorem{definition}{{{\textit{Definition}}}}

\DeclareMathOperator{\ngt}{neg}
\DeclareMathOperator{\rvr}{rvr}

\begin{document}

\title{A Computer Search of New OBZCPs of Lengths up to 49}

\author{Peter~Kazakov and Zilong~Liu
\thanks{Institute of Mathematics and Informatics, Bulgarian Academy of Sciences, Bulgaria (E-mail: {\tt peter.kazakov@math.bas.bg}). Zilong Liu is with School of Computer Science and Electronic Engineering, University of Essex, United Kingdom (e-mail: {\tt zilong.liu@essex.ac.uk}). }}

\maketitle

\begin{abstract}
This paper aims to search for new optimal and sub-optimal Odd Binary Z-Complimentary Pairs (OBZCPs) for lengths up to 49. As an alternative to the celebrated binary Golay complementary pairs, optimal OBZCPs are the best almost-complementary sequence pairs having odd lengths. We introduce a computer search algorithm with time complexity $O(2^N)$, where $N$ denotes the sequence length and then show optimal results for all $27 \le N \le 33$ and $N=37,41,49$. For those sequence lengths (i.e., $N=35,39,43,45,47$) with no optimal pairs, we show OBZCPs with largest zero-correlation zone (ZCZ) widths (i.e., $Z$-optimal). Finally, based on the Pursley--Sarwate criterion (PSC), we present a table of OBZCPs with smallest combined auto-correlation and cross-correlation.
\end{abstract}


\section{Introduction}\label{sec1}

\subsection{Background}
Complementary pairs of sequences are useful in coding theory, wireless communication, radar sensing, and signal processing. The general design objective is to find two equal-length sequences whose maximum aperiodic auto-correlation function (AACF) sums are as small as possible (ideally zero). Pioneered by Marcel J. E. Golay \cite{Golay-51} in 1951, binary \textit{complementary pairs} were first studied in his design of infrared multislit spectrometry, a detector which isolates the desired radiation with a fixed single wavelength from background radiation with many different wavelengths. Formally, a pair of sequences is called a Golay complementary pair (GCP) \cite{Golay-61} if their AACF sums are zero for all the non-zero time-shifts. Since it is  generally difficult to find a single unimodular sequence with zero AACF sidelobes\footnote{Although a Huffman sequence has zero AACF sidelobes except for the end time-shift, its sequence elements may have different magnitudes.}, GCP provides a solution by allowing two sequences to work in a collaborative way. Some representative applications of the GCPs (and their extensions/variants) include: peak-to-mean envelope power ratio (PMEPR) reduction of multicarrier signals \cite{popovic,DAVIS-JEWAB99,Zilong13,Zilong19}, Doppler resilient radar waveforms \cite{Turyn63,Doppler_GCP,Searle09}, channel estimation \cite{spasojevic,Wang07,Yuan08,Zilong20}, inter-cell interference rejection \cite{WHMow}, multicarrier code-division multiple access \cite{Liu-TCOM2014,Liu-TWC2015}, etc.

Despite their wide applications, however, binary GCPs are limited to even-lengths only. More specifically, the existing binary GCPs are only known to have sequence lengths of $2^\alpha 10^\beta 26^\gamma$ (where $\alpha,\beta,\gamma$ are non-negative integers). For the odd-length case, optimal binary almost--complementary pairs are studied in \cite{Zilong1}. Their optimization criteria is to maximize the zero-correlation zone (ZCZ) width of an odd-length binary pair, whilst minimizing its out-of-zone AACF sums. Such \textit{optimal} binary pairs exhibit the correlation properties closest to GCPs and are called \textit{optimal} odd-length binary Z-complementary pairs (OBZCPs). It is found that each optimal OBZCP has maximum ZCZ width of $(N+1)/2$, and minimum out-of-zone AACF sum magnitude of 2, where $N$ denotes the sequence length (odd) \cite{Zilong1}.

By applying insertion to certain binary Golay-Davis-Jedwab (GDJ) complementary pairs \cite{DAVIS-JEWAB99}, \textit{optimal} OBZCPs\footnote{In this paper, without any specific announcement, we are only interested in Type-I OBZCPs each having ZCZ around the in-phase time-shift. There are also Type-II OBZCPs where the ZCZs are centered around the end-shift positions, leading to zero correlation sidelobes away from the in-phase time-shift, but these are not our focus.} of lengths $2^m+1$, where $m$ is a positive integer, are constructed in \cite{Zilong1}. Subsequently, \textit{optimal} OBZCPs of \textit{generic} lengths  $2^\alpha 10^\beta 26^\gamma +1$ (where $\alpha,~ \beta, ~ \gamma$ are non-negative integers and $\alpha \geq 1$) are obtained in \cite{Zilong1b}, thanks to certain structural properties of binary GCPs obtained from Turyn's method \cite{Turyn74}.

\subsection{Contributions}
Searching for \textit{optimal} OBZCPs is of interest for understanding their deeper structural properties and for providing higher level of flexibility in practical applications. Besides the aforementioned systematic constructions, it is equally important to look for short OBZCPs because they may be used to generate longer OBZCPs through certain recursive operations. For example, in 2003, Borwein and Ferguson carried out an exhaustive computer search to search new GCPs of lengths up to 100 \cite{Borwein03}. Although some OBZCPs are reported in \cite{Zilong1}, their lengths are capped to 25 only.

The main objective of this paper is thus to look for new optimal (or sub-optimal) OBZCPs of lengths $27\leq N \leq 49$. By leveraging the High Performance Computing (HPC) facility\footnote{\url{https://www.essex.ac.uk/staff/it-services/hpc}.} at the University of Essex, a number of new OBZCPs are found by a fine-grained computer search algorithm involving mapping, grouping, and Gray code counting. Besides the OBZCP optimality criteria on maximum ZCZ width and minimum out-of-zone AACF sums, we also use the Pursley--Sarwate criterion \cite{Pursley1976,Katz0} for selection of the best binary odd-length pairs with lowest combined auto-correlation and cross-correlation. {For reproducibility of these results, our source code can be found at \url{https://github.com/peterkazakov/obzcp}}.

This paper is organized as follows. In Section II, we give some preliminaries and important equations/facts on OBZCPs. In Section III, we introduce the general considerations and programming techniques of the proposed algorithm. In Section IV, new OBZCPs found by computer search are presented. A list of OBZCPs with smallest smallest combined auto-correlation and cross-correlation are given in Section V. Finally, this paper is concluded in Section VI.

\section{Preliminaries}\label{sec2}
Throughout this paper, we are interested in binary sequence pairs whose elements are drawn from the set of $\mathbb{Z}_2=\{0,1\}$. A length-$N$ vector is called a binary sequence if it is over $\mathbb{Z}^N_2$. For convenience, whenever necessary, binary sequences may also be shown over $\{1,-1\}^N$. For $\mathbf{a}=(a_0,a_1,\cdots,a_{N-1})$ over $\mathbb{Z}^N_2$, let ${\mathbf{a}}(z)$ be the associated polynomial of $z$ as follows,
\begin{equation}\label{asso_poly_4_BiSeq}
{\mathbf{a}}(z)=\sum \limits_{\tau=0}^{N-1}(-1)^{a_\tau} z^{\tau}.
\end{equation}
For two binary sequences  ${\mathbf{{a}}}$ and $\mathbf{{{{{b}}}}}$ over $\mathbb{Z}^N_2$, define
\begin{equation}
\rho_{\mathbf{a},\mathbf{b}}(\tau)= \left \{
\begin{array}{cl}
\sum\limits_{i=0}^{N-1-\tau}(-1)^{a_i+b_{i+\tau}},&~~0\leq \tau \leq N-1;\\
\sum\limits_{i=0}^{N-1-\tau}(-1)^{a_{i+\tau}+b_{i}},&~~-(N-1)\leq \tau \leq -1;\\
0,& ~~|\tau|\geq N.
\end{array}
\right .
\end{equation}
When $\mathbf{a}\neq\mathbf{b}$, $\rho_{\mathbf{a},\mathbf{b}}(\tau)$ is called the aperiodic cross-correlation function of $\mathbf{a}$ and $\mathbf{b}$; otherwise, it is called the AACF. For simplicity, the AACF of ${\textbf{{a}}}$ will be sometimes written as $\rho_{\mathbf{a}}(\tau)$.

\vspace{0.1in}
\subsection{Binary Z-complementary pairs (ZCPs)}
\begin{definition}
Let $\mathbf{a}$ and $\mathbf{b}$ be over $\mathbb{Z}^N_2$. $(\mathbf{a},\mathbf{b})$ is said to be a binary ZCP with ZCZ width of $Z$ if and only if \cite{Fan07}
\begin{equation}\label{rho_eq}
\rho_{\mathbf{a}}(\tau)+\rho_{\mathbf{b}}(\tau)=0,~~\text{for any}~1\leq \tau \leq Z-1.
\end{equation}
In this case, $\rho_{\mathbf{a}}(\tau)+\rho_{\mathbf{b}}(\tau)$ for $Z\leq \tau\leq N-1$, is called the {out-of-zone aperiodic auto-correlation sum} of $\mathbf{a}$ and $\mathbf{b}$ at time-shift $\tau$. When $Z=N$, a ZCP reduces to a Golay complementary pair (GCP) \cite{Golay-61}. {An OBZCP refers to a binary ZCP with odd-length.}
\end{definition}
\vspace{0.1in}

\begin{fact}
Each OBZCP $(\mathbf{a},\mathbf{b})$ has the maximum ZCZ of width $(N+1)/2$ \cite{Fan11}, i.e., $Z\leq (N+1)/2$, where $N$ denotes the sequence length. An OBZCP is said to be $Z$-\textit{optimal} if $Z=(N+1)/2$.
\end{fact}

\vspace{0.1in}
\begin{fact}
The magnitude of each {out-of-zone aperiodic auto-correlation sum} for a $Z$-{optimal} OBZCP $(\mathbf{a},\mathbf{b})$ is lower bounded by 2 \cite{Zilong1}, i.e.,
\begin{displaymath}
\Bigl | \rho_{\mathbf{a}}(\tau)+\rho_{\mathbf{b}}(\tau) \Bigl |\geq 2,~~\text{for any}~(N+1)/2\leq \tau \leq N-1.
\end{displaymath}
A $Z$-\textit{optimal} OBZCP is said to be \textit{optimal} if $\Bigl | \rho_{\mathbf{a}}(\tau)+\rho_{\mathbf{b}}(\tau) \Bigl |=2$ holds for all $(N+1)/2\leq \tau \leq N-1$.
\end{fact}

\begin{fact}
For an {\textit{optimal}} (or a $Z$-\textit{optimal}) OBZCP $(\mathbf{a},\mathbf{b})$, the following equations are satisfied \cite{Zilong1}
\begin{equation}\label{pro1_equ4}
\left \{
\begin{array}{cl}
a_0+a_{N-1}+b_0+b_{N-1}     & \equiv ~0~(\text{mod}~2),\\
a_r+a_{N-1-r}+b_r+b_{N-1-r} & \equiv ~1~(\text{mod}~2),
\end{array}
\right.
\end{equation}
where $1\leq r \leq (N-3)/2$.
\end{fact}

\subsection{Pursley--Sarwate Criterion \cite{Pursley1976,Katz0}}
For a binary sequence pair $(\mathbf{a},\mathbf{b})$ of length $N$, the cross-correlation demerit factor between $\mathbf{a}$ and $\mathbf{b}$ is defined by
\begin{equation}
    \text{CDF}(\mathbf{a},\mathbf{b})=\frac{\sum\limits_{\tau=1-N}^{N-1}|\rho_{\mathbf{a},\mathbf{b}}(\tau)|^2}{N^2}.
\end{equation}

For a binary sequence $\mathbf{a}$, the auto-correlation demerit factor of $\mathbf{a}$ is defined by
\begin{equation}
    \text{ADF}(\mathbf{a})=-1+\text{CDF}(\mathbf{a},\mathbf{a}).
\end{equation}
Furthermore, let us define the Pursley--Sarwate criterion of binary pair $(\mathbf{a},\mathbf{b})$ as follows:
\begin{equation}
\label{PSC}
    \text{PSC}(\mathbf{a},\mathbf{b})=\sqrt{\text{ADF}(\mathbf{a})\cdot \text{ADF}(\mathbf{b})}+\text{CDF}(\mathbf{a},\mathbf{b}).
\end{equation}
According to Pursley and Sarwate \cite{Pursley1976}, we have $\text{PSC}(\mathbf{a},\mathbf{b})\geq 1$. Katz and Moore pointed out in \cite{Katz0} that binary GCPs of lengths $2^\alpha 10^\beta 26^\gamma$ (where $\alpha,\beta,\gamma$ are non-negative integers) satisfy the Pursley--Sarwate lower bound with equality, i.e., $\text{PSC}(\mathbf{a},\mathbf{b})=1$.

In this paper, we also use Pursley--Sarwate criterion as a sieve to select OBZCPs. Specifically, we look for \textit{optimal} OBZCPs or $Z$-\textit{optimal} OBZCPs with PSC values closest to 1.

\section{Proposed Algorithm}\label{sec4}

\subsection{General Considerations}
In this section we describe our main considerations for our proposed algorithm with the {time} complexity of $O(2^{N})$. Exhaustive computer search of all pairs is infeasible due to the high complexity of $O(2^{2N})$. We present a novel algorithm based on a two-step approach using the algebraic constructions and software data structures within a reasonable memory constraint.

In order to limit duplicated calculations, the first consideration is to exclude equivalent pairs. Similar to the definition in \cite{Fan07}, two pairs $(\mathbf{a}=(a_{0}, \ldots, a_{N-1}), \mathbf{b}=(b_{0}, \ldots, b_{N-1}))$ are said to be equivalent, if one can be obtained by the other with one of the following operations:
\begin{itemize}
\item Interchange: if $(\mathbf{a},\mathbf{b})$ is a solution, then so is $(\mathbf{b},\mathbf{a})$;
\item Negation: if $(\mathbf{a},\mathbf{b})$ is a solution, then so is $(\ngt(\mathbf{a}), \mathbf{b})$, where $\ngt(\mathbf{a})=(1-a_{0},1-a_1,\ldots,1-a_{N-1})$;
\item {Reversal}: if $(\mathbf{a},\mathbf{b})$ is a solution, then so is $(\rvr(\mathbf{a}), \mathbf{b})$, where $\rvr(\mathbf{a})=(a_{N-1},\ldots,a_1, a_0)$.
\end{itemize}
Naturally, a series of these operations above lead to an equivalent solution. For instance $(\ngt(\mathbf{b})$, $\rvr(\ngt(\mathbf{a})))$ is also equivalent to $(\mathbf{a},\mathbf{b})$.

To proceed, without loss of generality, we assume that  $a_{N-1}=b_{N-1}=1$. We then run as separate cases for the all possible combinations of the middle bits $a_{(N-1)/2}$, $b_{(N-1)/2}$ and $a_0=b_0$ due to  \eqref{pro1_equ4}.

To visualize, we consider the following OBZCP structure in our search algorithm:
{\small
\begin{displaymath}
\left [
\begin{matrix}
a_0 & \color{blue}{a_{1}} &  \color{blue}{\ldots} & \color{blue}{a_{(N-3)/2}} & \color{red}{a_{(N-1)/2}} & \color{blue}{a_{(N+1)/2}}  & \color{blue}{\ldots}  & \color{blue}{a_{N-2}} & 1 \\
a_0 & \color{blue}{b_1}  & \color{blue}{\ldots}  & \color{blue}{b_{(N-3)/2}}  & \color{red}{b_{(N-1)/2}}   & \color{blue}{b_{(N+1)/2}} & \color{blue}{\ldots} & \color{blue}{b_{N-2}} & 1
\end{matrix}
\right ].
\end{displaymath}
}

Next, we leverage the following three important data processing strategies in our proposed algorithm:
\begin{itemize}
\item Mapping;
\item Grouping;
\item Gray counting and updates.
\end{itemize}

\subsection{Mapping}

The purpose of mapping is to link each sequence $\mathbf{b}$ with a proper sequence $\mathbf{a}$ such that the ZCZ can be achieved.
Based on \eqref{rho_eq}, we have
\begin{equation}\label{reversed_rho}
\rho_{\mathbf{b}}(\tau)=-\rho_{\mathbf{a}}(\tau)
\end{equation}
for each $\tau=1,2,\ldots,(N-3)/2$. Suppose that there are $k$ sequences of $\mathbf{a}$ which satisfy the above equation. We store all the $\rho$ values and their corresponding sequences $\mathbf{a}$ in a map $\{ \rho_{\mathbf{a}} \rightarrow [\mathbf{a}_1,\ldots,\mathbf{a}_k]\}$ where
\[\rho_{\mathbf{a}}=(\rho_{\mathbf{a}}(1),  \rho_{\mathbf{a}}(2), \ldots, \rho_{\mathbf{a}}((N-3)/2) )
\]
for all $\mathbf{a} \in \{\mathbf{a}_1, \ldots, \mathbf{a}_k\}$ that generate the same $\rho$. {We will denote by $\text{map}[\rho_{\mathbf{a}}]$ for the whole list of $\{\mathbf{a}_1,\ldots,\mathbf{a}_k\}$}.


\begin{example}
For $N=11$, $\rho_{\mathbf{a}}=(2, -1, 2, 1)$, we have
\begin{displaymath}
\begin{aligned}
\text{map}[\rho_{\mathbf{a}}] = & \{(1,0,0,0,0,0,1,0,0,1,1),\\
                                                & ~ (1,0,1,1,0, 0,1,1,1,1,1),\\
                                                & ~ (1,1,0,0,1,0,0,0,0,0,1),\\
                                                & ~ (1,1,1,1,1,0,0,1,1,0,1)\}.
\end{aligned}
\end{displaymath}
\end{example}

For each sequence $\mathbf{b}$, proper software implementation of such map will permit us to find all possible corresponding sequences $\mathbf{a}$ in constant $O(1)$ time.

Theoretically, with this approach we can attack the whole problem. However, this may require us to put maximally all $2^N$ sequences of $\mathbf{a}$ and their corresponding $\rho$ values in the memory. This is quite significant considering the physical memory limitations. Next, we introduce grouping in order to overcome this limitation.

\subsection{Grouping}

In the first step, we group the sequence elements by using the property below:
\begin{equation}
a_r+a_{N-1-r}+b_r+b_{N-1-r}\equiv 1~(\text{mod}~2).
\end{equation}

{For ease of presentation, denote by $N_\mathbf{a}$ the integer associated to the binary sequence $\mathbf{a}$ and vice versa. }

{For any binary vector $\mathbf{c} = (c_1, c_2, \dotsc, c_{(N-3)/2)})$, define a set $\mathbf{C}$ comprising of all the possible length-$N$ binary sequences $\mathbf{a}$, each satisfying
\begin{equation}\label{chunk_def2}
c_i = a_i \oplus a_{N-1-i}, i=1, 2,\ldots, (N-3)/2
\end{equation}
where $\oplus $ is the binary xor operation. By noting that ${a}_0, {a}_{(N-1)/2}, {a}_{N-1}$ are fixed, each set $\mathbf{C}$ contains $2^{(N-3)/2}$ sequences of $\mathbf{a}$ and we have $2^{(N-3)/2}$ sets for all possible combinations of $\mathbf{c}$.
\begin{example}
Consider $N=7$, $Z=(N+1)/2=4$, $\mathbf{c}=(1,0)$. One can generate the following $\mathbf{C}{(\mathbf{c})}$: 
\begin{displaymath}
\begin{aligned}
& \{(a_0, 0, 0, a_3, 0, 1, 1), (a_0, 1, 0, a_3, 0, 0, 1),\\
& ~ (a_0, 0, 1, a_3, 1, 1, 1), (a_0, 1, 1, a_3, 1, 0, 1)\}.
\end{aligned}
\end{displaymath}
\end{example}
}

\subsection{Gray code counting}

For every sequence $\mathbf{b}$, note that $b_{N-1}=1$, $b_{(N-1)/2}$ is fixed and $b_0=a_0$ as shown in \eqref{pro1_equ4}. To proceed, $(b_1, b_2, \ldots, b_{(N-3)/2})$ is said to be the lower part of sequence $\mathbf{b}$ and $(b_{(N+1)/2}, b_{(N+3)/2}, \ldots, b_{(N-2)})$ its upper part. 
Our key idea is to use a Gray code to represent the lower part, whereby the upper part is recalculated based on \eqref{pro1_equ4}.

{
In the sequel, ``codeword" and ``sequence" may be used interchangeably. The weight of each codeword is equal to its number of 1's. Comparison between codewords (sequences), e.g., $\mathbf{a}\geq \mathbf{b}$, is performed by comparing their associated integers, i.e., $N_\mathbf{a}\geq N_\mathbf{b}$.
}
A Gray code is characterized by having the next codeword with only one bit changed. This gives us two advantages:

\begin{itemize}
\item We can update only one corresponding bit in $\mathbf{b}$'s upper part, i.e. if $b_i$ is changed, then $b_{N-1-i}$ is changed.
\item For such a change, the weight of sequence $\mathbf{b}$ is either increased or decreased by $2$ or stays unchanged. Hence, we can track the weight of sequence $\mathbf{b}$ without recalculating it for each modification. By pre-calculating all the admissible weight pairs associated to $(\mathbf{a},\mathbf{b})$ (see \textit{Property 3} of \cite{Zilong1}), we can directly exclude all the non-matching combinations.
\end{itemize}

\subsection{Sketch of the Proposed Algorithm}\label{sec7}



\begin{figure}[htp] \centering{
\includegraphics[scale=0.37]{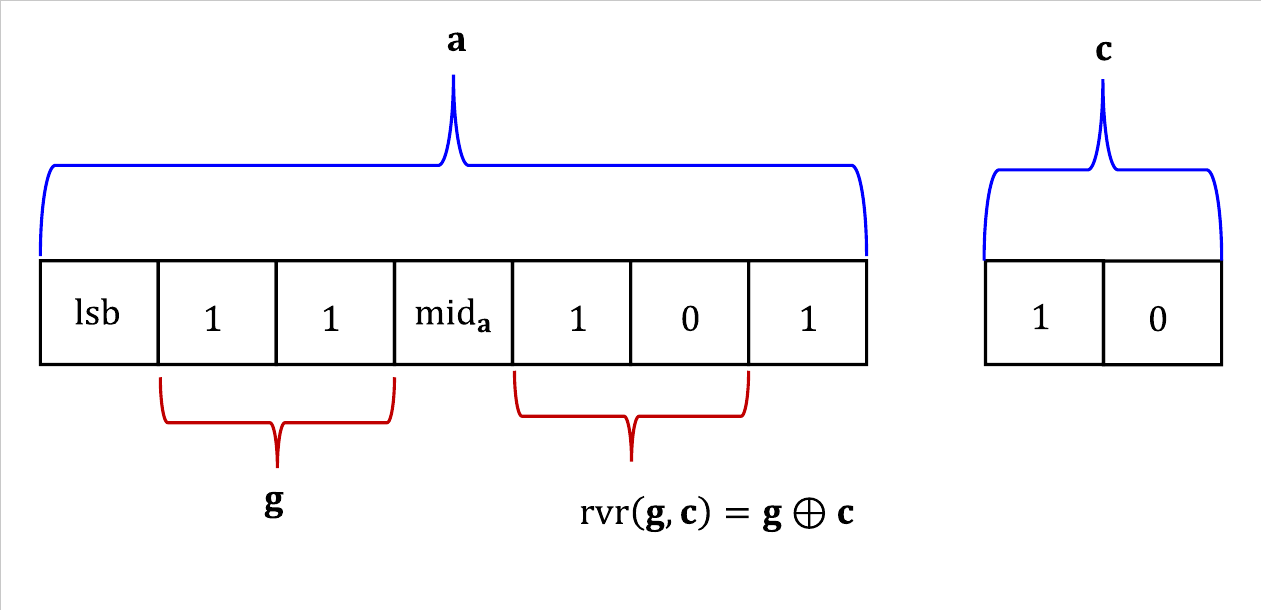}}
\caption{Structure of sequence $\mathbf{a}$ of length 7.}
\end{figure}

In the proposed algorithm, {$\rvr(\mathbf{g},\mathbf{c})$ is defined
as $\rvr(\mathbf{g} \oplus \mathbf{c})$, which is obtained by binary xor operation of $\mathbf{g}$ and $\mathbf{c}$, followed by reversing the resulting binary vector.}
 {Sequence $\mathbf{a}$ is constructed by
$(\text{lsb}, \mathbf{g}, \text{mid}_\mathbf{a}, \rvr(\mathbf{g},\mathbf{c}), 1)$, where $\text{lsb}$ and $\text{mid}_\mathbf{a}$ are single binary elements and $\mathbf{g}$, $\rvr(\mathbf{g},\mathbf{c})$ are binary vectors of length $(N-3)/2$. For illustration purpose, we show in Figure 1 a length-7 $\mathbf{a}$ which belongs to $\mathbf{C}$ for $\mathbf{g}=(1,1)$ and $\mathbf{c}=(1,0)$. In the algorithm, $\mathbf{b\_upper}$ and $\mathbf{b\_lower}$, two binary vectors of length $(N-3)/2$, are used for construction of $\mathbf{b}$.
}


\begin{algorithm}
\caption{OBZCPs}\label{algo1}
\begin{algorithmic}[1]
\Require $N \geq 5$, $N$ \text{is odd}
\Require $\text{MAX\_ACC} \ge 1 \text{, maximum correlation}$
{
\Require $\text{mid}_{\mathbf{a}}$, \text{predefined value for }$a_{(N-1)/2}$
\Require $\text{mid}_\mathbf{b}$, \text{predefined value for }$b_{(N-1)/2}$
\Require $\text{lsb}$, \text{least significant bit of }$\mathbf{a}$ \text{and} $\mathbf{b}$
}

\For{${N_{\mathbf{c}}} \gets 0$ to $2^{(N-3)/2}-1$}
    \For { $N_{\mathbf{g}} \gets 0$ \text{to} $2^{(N-3)/2}-1$} \algorithmiccomment{{Add $\mathbf{a} \in \mathbf{C}$ to the map}}
    \State $\mathbf{a} \gets (\text{lsb}, \mathbf{g}, \text{mid}_\mathbf{a}, {\rvr(\mathbf{g},\mathbf{c})}, 1)$
        \State Add $\mathbf{a}$ to $\text{map}[\rho_\mathbf{a}]$ \
    \EndFor
    \State $\mathbf{b\_lower} \gets (1, 0, 0, \ldots, 0)$ \algorithmiccomment{{Construct initial $\mathbf{b}$}}
    \State \text{Calculate} $\mathbf{b\_upper}$ \text{such that} Eq.\eqref{pro1_equ4} \text{holds for} $\mathbf{b}=(\text{lsb}, \mathbf{b\_lower}, \text{mid}_\mathbf{b}, \mathbf{b\_upper}, 1)$ \text{and given} $\mathbf{c}$\text{. { $\mathbf{a}$'s bits are already aggregated with an xor in ${\mathbf{C}}$}}.
    \For {$j \gets 1$ to $2^{(N-3)/2}$} \algorithmiccomment{{Iterate over possible $\mathbf{b}$ candidates via Gray code}}
    \State $\rho_{\mathbf{a}}(\tau) \gets -\rho_{\mathbf{b}}(\tau), \tau=1, \ldots, (N-3)/2 $ 
    \For {$\mathbf{a} \in \text{map}[\rho_\mathbf{a}]$}
    \If{ $N_{\mathbf{a}} > N_{\mathbf{b}}$} \algorithmiccomment{{Skip duplication}}
    \State \text{Print} $(\mathbf{a},\mathbf{b})$ \text{if out-of-zone correlation} $\le \text{MAX\_ACC}$
    \EndIf
    \EndFor
    \State \text{Generate next Gray code for~}$\mathbf{b\_lower}$ \text{~part by changing single bit} $\mathbf{b}_x$
    \State \text{Update~}$\mathbf{b\_upper}$\text{~part by {reverting single } bit~ }\\$\mathbf{b}_{N-1-x}$
    \EndFor
\EndFor

\State Filter out { the} equivalent pairs

\end{algorithmic}
\end{algorithm}

The proposed algorithm is based on one outer loop to iterate over all the sets and two consecutive inner loops for 1) loading data in the memory map and 2) iterating over {the} Gray code by changing a single bit on each step and retrieving matching values from the map such that equation \eqref{pro1_equ4} holds.

{It is noted that the time complexity of the outer loops equals to the number of sets that we iterate on, i.e., $O(2^{(N-3)/2})$.
The time complexity of the two consecutive inner loops is determined by the maximum of:

\begin{itemize}
\item The complexity of the first inner loop which equals to the number of all constructed $\mathbf{a}$'s in each set, i.e., $O(2^{(N-3)/2)}$;
\item The complexity of the second inner loop which equals to the number of all the $\mathbf{b\_upper}$ values that we iterate on via Gray code, i.e., $O(2^{(N-3)/2})$. During the second inner loop, we construct all the possible $\mathbf{b}$ matching to the set $\mathbf{C}(\mathbf{c})$. It is also noted that the corresponding search in the table between $\mathbf{a}$ and $\mathbf{b}$ has lookup complexity of $O(1)$.
{\color{black}
Additionally, the most inner loop on line 10 is usually of size $0, 1, 2,$ and rarely up to $4$ and hence does not impact the overall complexity.}
\end{itemize}

Thus, the total time complexity of the inner loops is $O(2^{(N-3)/2})$.}

Based on the above analysis, the {overall} algorithm has a complexity of $O(2^N)$ {since it is executed } 8 times with all possible combinations of $a_{(N-1)/2},b_{(N-1)/2}, a_0$.
\color{black}


\section{Computer Search Results}\label{sec5}

In this section, we present optimal OBZCPs from computer search, and in case they do not exist, $Z$-optimal OBZCPs for $27 \le N \le 49$, in Tables I to XII.

Optimal OBZCPs up to $N=25$ are reported in \cite{Zilong1} and some optimal pairs for $N=33, 41$ can be obtained via the constructions in \cite{Zilong1b}. During this computer search, we have found only two non-equivalent optimal pairs for $N=37$. We have also found optimal pairs of length $N=49$, but the search is non-exhaustive as we reached the limit of our computational resources.

Our computer search shows that there are no optimal OBZCPs for $N=35, 39, 43, 45, 47$ and this motivates us to search $Z$-optimal pairs with the maximum out-of-zone aperiodic autocorrelation sum having magnitude of $6$.
In this work, we only present a few $Z$-optimal pairs in the tables below for $N=35, 39, 43, 45, 47$.

It is noted that all the OBZCPs are presented in hexadecimal notation. This allows us to map every hexadecimal digit to four binary digits, i.e. ($0 \rightarrow 0000$, $1 \rightarrow 0001$, \ldots, $F \rightarrow 1111$) and by stripping initial zeros if necessary. For instance {\tt 159FE24} represents $1010110011111111000100100$.

\begin{table*}[!ht]
\ttfamily
\centering
\caption{Optimal OBZCPs for $N=27$}\label{tab1}%
\begin{tabular}{@{}llll@{}}

\toprule
Column 1 & Column 2  & Column 3 & Column 4\\
\midrule

(6AC2984, 42265F0) &
(419B094, 4038DAA) &
(72C2320, 6581DAA) &
(5CB287E, 409159C) \\
(6A92700, 41D3994) &
(77724F1, 71B4ABF) &
(623AFC9, 4C14A0D) &
(668A84F, 4F0B77B)\\
\hline
\end{tabular}
\end{table*}

\begin{table*}[!ht]
\ttfamily
\centering
\caption{Optimal OBZCPs for $N=29$}\label{tab1}%
\begin{tabular}{@{}lll@{}}
\toprule
Column 1 & Column 2  & Column 3 \\
\midrule
(144E4E10, 114693FA) &
(17CA7B3A, 14640A7C) &
(1FCADB38, 1C64AA7E) \\
(1835D190, 17CA5190) &
(1A9FDB38, 15605B38) &
(1F6CC570, 11D1AD20) \\
(1AEF4E13, 16F06EEB) \\
\hline
\end{tabular}
\end{table*}

\begin{table*}[!ht]
\ttfamily
\centering
\caption{Optimal OBZCPs for $N=31$}\label{tab1}%
\begin{tabular}{@{}lll@{}}
\toprule
Column 1 & Column 2  & Column 3 \\
\midrule
(4ED3AE80, 43856426) &
(6C7FC945, 5EB32F1D) &
(64AFE6B9, 640ACBC7) \\
\hline
\end{tabular}
\end{table*}

\begin{table*}[!ht]
\ttfamily
\centering
\caption{Optimal OBZCPs for $N=33$}\label{tab1}%
\begin{tabular}{@{}llll@{}}
\toprule
Column 1 & Column 2  & Column 3\\
\midrule
(1EA6D8C0A, 1EA6C73F4) &
(180CBE6AC, 180CA1952) &
(1F8B39ED4, 1F8B2612A) \\
(1E72A06CA, 1A6C05630) \\
\hline
\end{tabular}
\end{table*}

\begin{table*}[!ht]
\ttfamily
\centering
\caption{$Z$-optimal OBZCPs for $N=35$}\label{tab1}%
\begin{tabular}{@{}lll@{}}
\toprule
Column 1 & Column 2  & Column 3 \\
\midrule
(7905A9444, 710C1A3B2) &
(72E2F6394, 5DE937410) &
(4AC870914, 40DDE22C2) \\
(54C870928, 40DDE22C2) &
(793B3EE96, 5C6F5043A) &
(44EA20C2D, 42D04DDE3) \\
\hline
\end{tabular}
\end{table*}

Since $N=35$ is the first odd length that does not have optimal pairs, we present on the left-hand-side of Fig. 1 on the magnitude plot of the AACF sums of the pair {\tt (7905A9444, 710C1A3B2)}. On the right-hand-side of Fig. 1, we show the magnitude plot of the optimal pair {\tt (15BCD1FAF3340,10599EA0E984A)} with length 49.

\begin{table*}[!ht]
\ttfamily
\centering
\caption{Optimal OBZCPs for $N=37$}\label{tab1}%
\begin{tabular}{@{}ll@{}}
\toprule
Column 1 & Column 2 \\
\midrule
(1D29F4D110, 11273940E8) &
(17B506C9C4, 144430A7C2) \\
\hline
\end{tabular}
\end{table*}

\begin{table*}[!ht]
\ttfamily
\centering
\caption{$Z$-optimal OBZCPs for $N=39$}\label{tab1}%
\begin{tabular}{@{}lll@{}}
\toprule
Column 1 & Column 2  & Column 3 \\
\midrule
(69B1294470, 5AEF8C06E8) &
(6CC43E1164, 4A75EC2828) &
(70122A279C, 5B235A9E08) \\
(72577E7298, 6E0592287A) &
(59F26A2072, 44AF3882D0) &
(78C84D1254, 77E235C7D2)\\
\hline
\end{tabular}
\end{table*}

\begin{table*}[!ht]
\ttfamily
\centering
\caption{Optimal OBZCPs for $N=41$}\label{tab1}%
\begin{tabular}{@{}lll@{}}
\toprule
Column 1 & Column 2  & Column 3 \\
\midrule
(1A2903A133C, 15D6FCA133C) &
(18215995906, 1821586A6F8) &
(18A5F99F9AE, 175A069F9AE) \\
(1945F99F9D6, 16BA069F9D6) &
(1F5EC9C62FA, 10A136C62FA) &
(195DC396EFC, 16A23C96EFC) \\
(1F51C9C6DFA, 10AE36C6DFA) &
(1A2903C337A, 15D6FCC337A) &
(1E14DDB4188, 11EB22B4188) \\
(1C9EF59CBA0, 13610A9CBA0) &
(1A774FA7BB0, 1588B0A7BB0) &
(19FA296B9F9, 1605D66B9F9) \\

\hline
\end{tabular}
\end{table*}

\begin{table*}[!ht]
\ttfamily
\centering
\caption{$Z$-optimal OBZCPs for $N=43$}\label{tab1}%
\begin{tabular}{@{}lll@{}}
\toprule
Column 1 & Column 2  & Column 3 \\
\midrule
(668D847F75A,62EEF0F6BB4), &
(74BF732EE12,43637D50C1A), &
(7121C0324F5,4AB9D90D7E5) \\
\hline
\end{tabular}
\end{table*}

\begin{table*}[!ht]
\ttfamily
\centering
\caption{$Z$-optimal OBZCPs for $N=45$}\label{tab1}%
\begin{tabular}{@{}lll@{}}
\toprule
Column 1 & Column 2  & Column 3 \\
\midrule
(1C3009533530,15C8AF20693C) &
(1ADB163A02A9,118C8F16C00D) &
(13589BC02838,11EF69162A6E) \\
\hline
\end{tabular}
\end{table*}

\begin{table*}[!ht]
\ttfamily
\centering
\caption{$Z$-optimal OBZCPs for $N=47$}\label{tab1}%
\begin{tabular}{@{}lll@{}}
\toprule
Column 1 & Column 2  & Column 3 \\
\midrule
(5D9F943CC229,4BBA9D0B6FE3) &
(5375CCE48E1E,4752A49003F4) &
(7815ACEB273E,579B7779603A) \\
\hline
\end{tabular}
\end{table*}

\begin{table*}[!ht]
\ttfamily
\centering
\caption{\textit{Optimal} OBZCPs for $N=49$}\label{tab1}%
\begin{tabular}{@{}ll@{}}
\toprule
Column 1 & Column 2  \\

\midrule
(15BCD1FAF3340,10599EA0E984A) &
(1FB67150F994A,1A533E0AE3240) \\
(1564E7FF86694,152CC3E031B2A) &
(156CC7FF8E494,1524E3E03992A) \\
(15BCD1FFA6614,150CCBE0E984A)&
(1FB67155ACC1E,1F066B4AE3240)\\
(1564E7FAD33C0,107996A031B2A)&
(156CC7FADB1C0,1071B6A03992A)\\
\hline
\end{tabular}
\end{table*}


\begin{figure*}[ht]
\begin{minipage}[b]{0.45\linewidth}
\centering
\includegraphics[trim=2.5cm 8.5cm 2.5cm 8.5cm, clip=true, scale = 0.6]{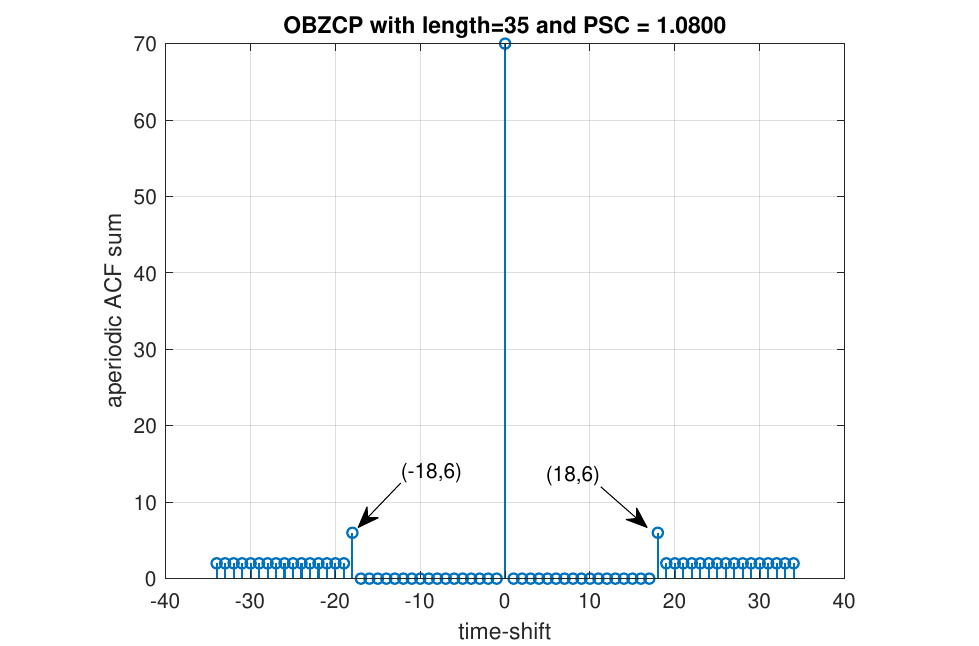}
\label{fig:figure1}
\end{minipage}
\hspace{0.6cm}
\begin{minipage}[b]{0.45\linewidth}
\centering
\includegraphics[trim=2.5cm 8.5cm 2.5cm 8.5cm, clip=true, scale = 0.6]{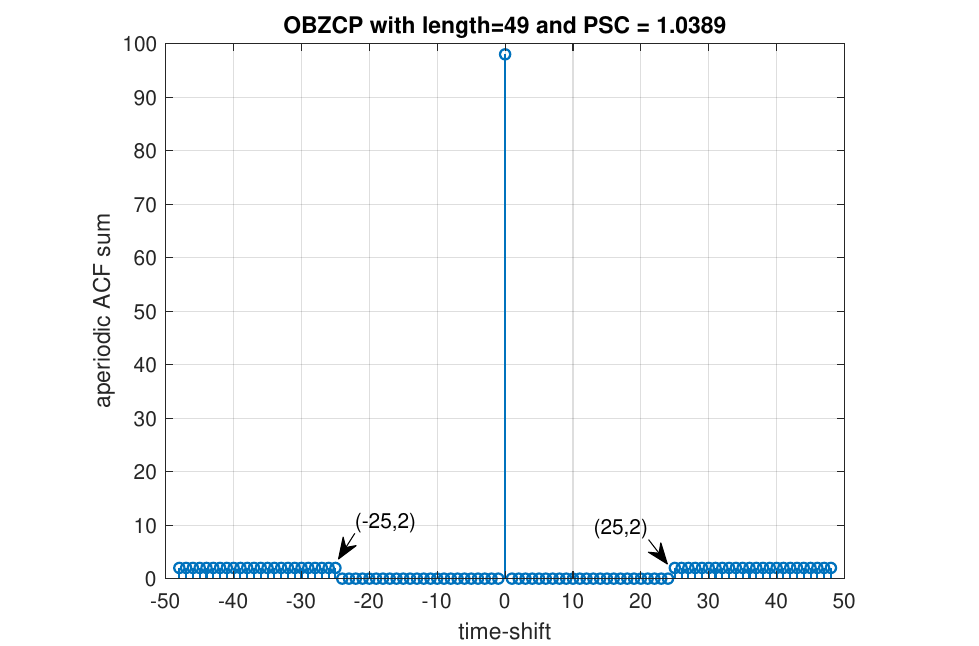}
\label{fig:figure2}
\end{minipage}
\caption{Plots of AACF sums for OBZCPs, i.e., {\tt (7905A9444, 710C1A3B2)} and {\tt (15BCD1FAF3340,10599EA0E984A)}, of lengths $35$ and $49$, respectively.}
\end{figure*}



{
Our algorithm is implemented using the software language ``Rust" known as blazingly fast and memory-efficient \cite{Rust1} and includes optimizations such as Rayon parallel computing library, usage of native CPU instructions via compilation flags, such as SIMD, loops unwinding, etc. Our source code can be found at \url{https://github.com/peterkazakov/obzcp}.
The OBZCP search for lengths up to 41 was carried out over a 2.9 GHz 6-Core Intel i9 computer. The execution time over for $N=31$ and $N=33$ are $3$ and $14$ minutes, respectively.
Each subsequent length requires approximately $4$ times more minutes, for instance $N=39$ requires a day of computations.

The other results for $N=43,45,47,49$ are found through Essex HPC.}

\section{OBZCPs with Smallest PSC Values}

Finally, we leverage the Pursley--Sarwate criterion and present OBZCPs with smallest PSC values for sequence lengths up to 49 in Table XIII. 

\begin{table*}[!ht]
\ttfamily
\centering
\caption{OBZCPs with smallest PSC values}\label{tab1}%
\begin{tabular}{@{}lllllll@{}}
\toprule
N &
PSC($g_1$,$g_2$) &
ADF($g_1$) &
ADF($g_2$) &
CDF($g_1$,$g_2$) &
$g_1$ &
$g_2$ \\

\hline
3 & 1.4444 & 1.1111 & 1.1111 & 0.33333 & 7 & 5\\
5 & 1.2997&0.48& 0.8&0.68& 1E&16\\
7 & 1.185 & 0.77551 & 0.28571 & 0.71429 & 5D & 4F\\
9 & 1.1636 & 0.79012 & 0.39506 & 0.60494 & 1E8 & 14C\\
11 & 1.149  & 0.41322 & 0.67769 & 0.61983 & 7C6 & 5DA \\
13 & 1.1005 & 0.07100 & 0.30769 & 0.95266 & 1F35 & 1709 \\
15 & 1.1107 & 0.38222 & 0.20444  & 0.83111 & 612E & 4C0A \\

17 & 1.0983 & 0.38754 & 0.609 & 0.61246 &  1FCA9 & 1A6E3 \\

19 & 1.0931 & 0.40443 & 0.27147 & 0.76177 & 7A284 &  78CDA \\

21 &
1.0819 & 0.29932 & 0.46259 & 0.70975 & 1B6A87 & 109883 \\

23 & 1.0762 & 0.41966 & 0.58601 & 0.58034 & 78B519 & 76D1DF \\

25 &
1.0740 &
0.4224 &
0.5248 &
0.6032 &
159FB70 &
11DA0CA \\

27&
1.0664 &
0.36488 &
0.25514 &
0.76132 &
5CB287E &
409159C\\

29 &
1.0608 &
0.27111 &
0.39477 &
0.73365 &
144E4E10 &
114693FA\\

31 &
1.0571 &
0.4308 &
0.30593 &
0.69407 &
64AFE6B9 &
640ACBC7 \\

33 &
1.0554 & 0.33792 & 0.24977 & 0.76492 & 1E72A06CA & 1A6C05630 \\

35 &
1.0792 &
0.19102 & 0.25633 &
0.85796 &
44EA20C2D &
42D04DDE3 \\

37 &
1.0515 &
0.30095 &
0.35354 &
0.72535 &
1D29F4D110 &
11273940E8 \\

39 &
1.0874 & 0.27745 & 0.3879 & 0.75937 & 78C84D1254 & 77E235C7D2 \\

41 &
1.0460 & 0.38548 & 0.31886 &  0.69542 & 19FA296B9F9 & 1605D66B9F9 \\

43 &
1.0964 & 0.38615 &  0.33423 & 0.73716 &
668D847F75A &
62EEF0F6BB4 \\

45 &
1.0748 & 0.37728 & 0.34963 & 0.7116&
1ADB163A02A9&
118C8F16C00D \\

47 &
1.0972&0.32866& 0.41195& 0.72929& 5D9F943CC229& 4BBA9D0B6FE3\\

49 &
1.0388 &
0.33986 &
0.39983 &
0.67014 &
156CC7FF8E494 &
1524E3E03992A \\

\hline
\end{tabular}
\end{table*}

One can see that  the PSC value of each pair is close to $1$, indicating that each pair is almost complementary. We also notice a trend that the PSC values for optimal OBZCPs generally decrease for larger sequence lengths with just one exception for $N=13$.

\section{Conclusions}\label{sec13}
 This paper has introduced a number of new primitive \textit{optimal} and $Z$-\textit{optimal} OBZCPs for $27\leq N \leq 49$ which are obtained through a computer search.  By using selected algebraic properties of OBZCPs and with certain programming techniques, we have presented an algorithm with time complexity of $O(2^N)$. Next to the known OBZCP constructions for $N=33$ and $N=41$, we have found \textit{optimal} sequence pairs for $N=37$ and $N=49$. Additionally, a list of OBZCPs with smallest PSC values have been reported.

 An interesting future work of this research is to develop systematic \textit{optimal} (or sub-\textit{optimal}) OBZCP constructions by exploiting the obtained \textit{short} sequence pairs. Another ambitious future direction is to carry out an exhaustive computer search for OBZCPs up to length 100 and analyze their deeper structural properties.


\end{document}